# Lower bounds on the rate-distortion function of LDGM codes

A. G. Dimakis[1]   M. J. Wainwright[1,2], and   K. Ramchandran[1]

[1]Department of Electrical Engineering and Computer Science
[2]Department of Statistics
UC Berkeley, Berkeley, CA.

## Abstract

A recent line of work has focused on the use of low-density generator matrix (LDGM) codes for lossy source coding. In this paper, we develop a generic technique for deriving lower bounds on the rate-distortion functions of binary linear codes, with particular interest on the effect of bounded degrees. The underlying ideas can be viewing as the source coding analog of the classical result of Gallager, providing bounds for channel coding over the binary symmetric channel using bounded degree LDPC codes. We illustrate this method for different random ensembles of LDGM codes, including the check-regular ensemble and bit-check-regular ensembles, by deriving explicit lower bounds on their rate-distortion performance as a function of the degrees.

## 1 Introduction

The problem of lossy source coding is to achieve maximal compression of a data source, subject to some bound on the average distortion. Classical random coding arguments show that a randomly chosen binary linear code will, with high probability, come arbitrarily close to the rate-distortion bound for lossy compression of a symmetric Bernoulli source [4]. However, such codes are impractical, as it is neither possible to represent them in an compact manner, nor to perform encoding/decoding in an efficient way. It is thus of considerable interest to explore and analyze the use of structured codes for lossy compression. A particularly important subclass of structured codes are those based on bounded-degree graphs, such as trellis codes, low-density parity check (LDPC) codes, and low-density generator matrix (LDGM) codes. One practical approach to lossy compression is via trellis-code quantization [13]. One limitation of trellis-based approaches is the fact that saturating rate-distortion bounds requires increasing the trellis constraint length [22], which incurs exponential complexity (even for the max-product or sum-product message-passing algorithms). Other work shows that it is possible to approach the binary rate-distortion bound using LDPC-like codes [17] or nonlinear codes [11], albeit with degrees that grow at least logarithmically with the blocklength. A parallel line of recent work [16, 23, 1, 2, 21] has explored the use of low-density generator matrix (LDGM) codes for lossy compression. These codes correspond to the duals of low-density parity check (LDPC) codes, and thus can be represented in terms of sparse factor graphs. The results of this paper provide further insight into the effective rate-distortion function of this class of sparse graph codes.

Focusing on binary erasure quantization (a special compression problem dual to binary erasure channel coding), Martinian and Yedidia [16] proved that LDGM codes combined with modified message-passing can saturate the associated rate-distortion bound. Various researchers have used techniques from statistical physics, including the cavity method and replica methods, to provide non-rigorous analyses of LDGM performance for lossy compression of binary sources [1, 2, 21]. In the limit of zero-distortion, this analysis has been made rigorous in a sequence of papers [5, 19, 3, 7]. The papers [15, 14] provide rigorous upper bounds on the effective rate-distortion function of various classes of LDGM codes, assuming the use of a maximum likelihood (ML) encoder. In terms of practical algorithms for lossy binary compression, several researchers have explored variants of the sum-product algorithm or survey propagation algorithms [1, 8, 21, 23] for quantizing binary sources.

Previous rigorous analyses of the effective rate-distortion function of LDGM codes [15, 14] under ML encoding have been based on the first and second-moment methods. Whereas the second moment provides a non-trivial upper bound on the effective rate-distortion function, the first moment method (at least in its straightforward application) yields a well-known statement—namely, that the rate must be larger than than the Shannon rate-distortion. This lower bound, though achieved for graphs with degrees that scale suitably with blocklength, is far from sharp for these sparse graph codes. Accordingly, the primary contribution of this paper is the development of a technique for generating sharper *lower bounds* on the effective rate-distortion function of sparse graph codes. At a high-level, the core of our approach can be understood as a source coding analog of Gallager's [9] classical result on the effective capacity of bounded degree LDPC codes for channel coding. Our main result (Theorem 1) shows explicitly how, for fixed sequences of codes, the gap to the rate-distortion bound is controlled by a certain measure of the *average overlap* between quantization balls. We illustrate our approach in application to some random ensembles of LDGM codes, establishing how their effective rate-distortion performance compares to the Shannon limit for various bit and check degrees. We note that since this work was initially presented [6], Kudekar and Urbanke [12] have used related methods to establish lower bounds that hold for fixed codes, as opposed to the random ensemble analysis of this paper.

The remainder of this paper is organized as follows. We begin in Section 2 with necessary background material and definitions for source coding, factor graphs, and low-density generator matrix codes, before stating and discussing our main results in Section 2.3. Section 3 is devoted to a number of basic results, applicable to any binary linear code. These results show how lower bounds on the effective rate-distortion can be obtained by suitably lower bounding the growth rate of the number of codewords with sufficiently low distortion. In general, this growth rate—represented as a certain average of the overlaps between codewords—is a complicated quantity to analyze, due to the non-uniform nature of the underlying random variable. Nonetheless, as we show in Section 4, it is possible to obtain explicit and computable lower bounds on the average-case performance of LDGM codes, using a graph-based certificate and ensemble averages to obtain explicit lower bounds on the relevant overlap, and hence on the rate-distortion function. This work was first presented in part at the Information Theory Workshop, Lake Tahoe [6].

## 2 Background and Statement of Main Results

We begin with background on binary linear codes, lossy source coding, factor graphs, and random ensembles of low-density generator matrix codes. With these definitions, we then state our main results in Section 2.3.

### 2.1 Binary codes and lossy source coding

A binary linear code $\mathbb{C}$ of block length $n$ consists of a linear subspace of $\{0,1\}^n$. One concrete representation is as the range space of a given generator matrix $\mathbf{G} \in \{0,1\}^{n \times m}$, as follows:

$$\mathbb{C} = \{x \in \{0,1\}^n \mid x = \mathbf{G}z \quad \text{for some } z \in \{0,1\}^m \}. \tag{1}$$

The code $\mathbb{C}$ consists of at most $2^m = 2^{nR}$ codewords, where $R = \frac{m}{n}$ is the code rate.

In the binary lossy source coding problem, the encoder observes a symmetric Bernoulli source sequence $S \in \{0,1\}^n$, with each element $S_i$ drawn in an independent and identically distributed (i.i.d.) manner from a Bernoulli distribution with parameter $p = \frac{1}{2}$. The idea is to compress the source by representing each source sequence $S$ by some codeword $x \in \mathbb{C}$. When using a code in generator matrix form, one thinks of mapping each source sequence to some codeword $x \in \mathbb{C}$ from a code containing $2^m = 2^{nR}$ elements, say indexed by the binary sequences $z \in \{0,1\}^m$. The source decoding map $x \mapsto \widehat{S}(x)$ associates a source reconstruction $\widehat{S}(x)$ with each codeword $x \in \mathbb{C}$. The quality of the reconstruction can be measured in terms of the Hamming distortion $d(S, \widehat{S}) = \sum_{i=1}^{n} |S_i - \widehat{S}_i| = \|S - \widehat{S}\|_1$. With this set-up, the source encoding problem is to find the codeword with minimal distortion—namely, the optimal encoding $\widehat{x}_{ML} := \arg\min_{x \in \mathbb{C}} d(\widehat{S}(x), S)$. Classical rate-distortion theory [4] dictates that, for the binary symmetric source, the optimal trade-off between the compression rate $R$ and the best achievable average distortion $D = \mathbb{E}[d(\widehat{S}, S)]$ is given by

$$R(D) = 1 - H(D), \tag{2}$$

where $H(D) := -D \log D - (1-D) \log(1-D)$ is the binary entropy function.

## 2.2 Factor graphs, LDGM codes and random ensembles

Given a binary linear code $\mathbb{C}$, specified by generator matrix $\mathbf{G}$, the code structure can be captured by a bipartite graph, in which square nodes (■) represent the checks attached to the code output bits $x_i$ (or rows of $\mathbf{G}$), and circular nodes (○) represent the information bits (or columns of $\mathbf{G}$). For instance, Fig. 1 shows the factor graph for a rate $R = \frac{3}{4}$ code in generator matrix form, with $n = 12$ checks (each associated with a unique source bit, top of diagram) connected to a total of $m = 9$ information bits (bottom of diagram). The edges in this graph correspond to 1's in generator matrix matrix, and reveal the subset of bits to which each information bit contributes. The degrees of the check (respectively) variable nodes in the factor graph are $d_c = 3$ and $d_v = 4$ respectively, so that the associated generator matrix $\mathbf{G}$ has 3 ones in each row, and 4 ones in each column. When the generator matrix is sparse, then the resulting code is known as a low-density generator matrix (LDGM) code. Some authors also refer to codes in which the degrees scale sublinearly with blocklength as low-density; in this paper, we reserve this term only for codes with degrees bounded independently of the blocklength.

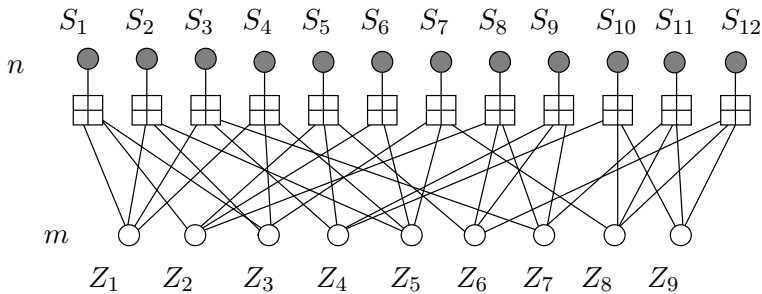

**Figure 1.** Factor graph representation of an LDGM code with $n = 12$ source bits, $m = 9$ checks, and overall rate $R = \frac{m}{n} = \frac{3}{4}$. The code illustrated is a bit-check-regular code from the ensemble $\mathfrak{C}(d_c, d_v)$, with bit degree $d_v = 4$ and check degree $d_c = 3$.

Our primary contribution is a technique for generating lower bounds on the rate-distortion functions of binary linear codes. We illustrate this method concretely by application to two different random ensembles of LDGM codes. The *check-regular LDGM ensemble with degree $d_c$*, denoted by $\mathfrak{C}(d_c)$, is formed by fixing a check degree $d_c$, and having each of the $n$ checks connect to $d_c$ of the $m$ information bits uniformly at random (with replacement). Doing so generates a set of information bits with a random degree sequence, one which asymptotically obeys a Poisson law with mean $d_c/R$. This particular ensemble of random graphs is the canonical choice in studying random $k$-SAT, XORSAT and other satisfiability problems (e.g., [18, 5, 7]). Note that the problem of LDGM encoding—finding the sequence of information bits to minimize Hamming distortion—is equivalent to an instance of a MAX-XORSAT problem.

On the other hand, the *bit-check-regular LDGM ensemble*, denoted by $\mathfrak{C}(d_c, dv)$, is specified by a pair of degrees $(d_c, d_v)$, one for the checks and one for the bits. The code ensemble consists of all codes in each each check has degree exactly $d_c$ and each bit has degree exactly $d_v$. For example, the code illustrated in Figure 1 is bit-check-regular with $(d_c, d_v) = (3, 4)$. This ensemble is the LDGM analog of the Gallager regular ensembles [9] of LDPC codes.

### 2.3 Main results

Given a $\mathbb{C}$, define for each binary string $u \in \{0, 1\}^n$, the integer-valued quantity

$$N(u, D; \mathbb{C}) := \left|\{z \in \{0, 1\}^m \mid \mathbf{G}z \in \mathbb{B}_n(u; D)\}\right|, \qquad (3)$$

that counts the number of information sequences that generate codewords within the Hamming $D$-ball $\mathbb{B}_n(s; D)$ centered at $s$. We let $\mathbb{E}_U[1/N(U, D; \mathbb{C})]$ denote the average of $1/N$, where $U$ is uniformly distributed over the ball $\mathbb{B}_n(\vec{0}; D)$. With these definitions, we have

**Theorem 1.** *Consider a fixed sequence of codes $\mathbb{C} \equiv \mathbb{C}_n$ of rate $R$, indexed by blocklength $n$. If for sufficiently large $n$, the rate-distortion pair $(R, D)$ satisfies the bound*

$$R < 1 - H(D) - \frac{1}{n} \log \mathbb{E}_U[1/N(U, D; \mathbb{C})], \qquad (4)$$

*then the code family cannot achieve distortion $D$.*

Since by definition, we have $N(u, D; \mathbb{C}) \geq 1$ for any $u \in \mathbb{B}_n(\vec{0}; D)$, we always have the trivial lower bound $-\log \mathbb{E}_U[1/N(U, D; \mathbb{C})] \geq 0$, under which the bound (4) reduces to the Shannon rate-distortion bound. Indeed, this bound would be asymptotically tight for a random (high-density) linear code. For other codes, obtaining more refined statements requires exploiting specific aspects of the code structure.

Theorem 1 holds for any fixed (deterministic) sequence of codes. In order to use it to establish lower bounds on the rate-distortion function of given code sequences. one needs upper bounds on the quantity $\frac{1}{n} \log \mathbb{E}_U[1/N(U, D; \mathbb{C})]$. The analysis of this quantity is facilitated by considering random ensembles of codes. In particular, in Section 4, we analyze the behavior of the lower bound (4) for different random ensembles of codes, thereby obtaining explicit lower bounds as corollaries of Theorem 1. We begin with the check-regular ensemble:

**Corollary 1.** *With probability converging to one with the blocklength, a LDGM code $\mathbb{C}(d_c)$ randomly drawn from the check-regular ensemble $\mathfrak{C}(d_c)$ with check degree $d_c$ can only achieve those rate-distortion pairs $(R, D)$ that satisfy the bound*

$$R\left[1 - \frac{1}{2}\exp(-\frac{(1-D)d_c}{R})\right] \geq 1 - H(D). \qquad (5)$$

*For any finite degree $d_c$, the minimal rate $R$ satisfying the relation (5) is strictly bounded away from the Shannon rate-distortion bound $R(D) = 1 - H(D)$.*

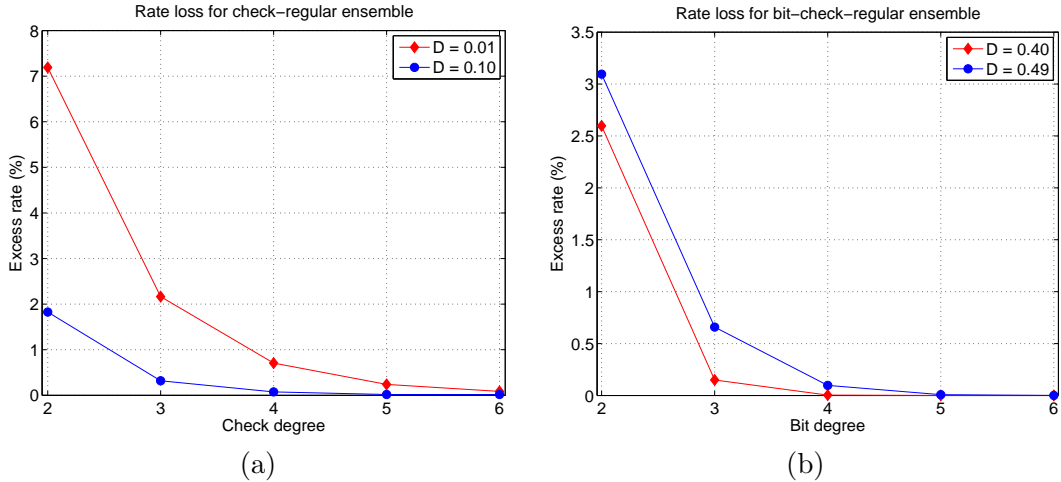

**Figure 2.** Excess rate, computed as a percentage of the Shannon limit $R(D) = 1 - H(D)$, versus bit/check degrees for different random ensembles. (a) Plot of the lower bound from Corollary 1 for the check-regular ensemble as a function of the check degree $d_c$. (b) Plot of the lower bound from Corollary 2 for the bit-check-regular ensemble as a function of the bit degree $d_v$.

Figure 2(a) plots the minimal rates $R$ satisfying the bound (5), as a function of the check degree $d_c$, for two different distortions. Although Corollary 1 bounds the rate-distortion function away from the Shannon limit for any finite degree, note that the performance loss decreases rapidly as the check degree $d_c$ is increased. However, we do know that the bound (5) is not sharp (in particular, it is loose in the special case $D = 0$), and we suspect that our analysis could be refined in a a number of places.

We also have a complementary result for the case of bit-check-regular ensembles:

**Corollary 2** (Bit-check-regular ensembles). *With probability converging to one with the blocklength, a LDGM code $\mathbb{C}(d_c, d_v)$ randomly drawn from the bit-check-regular ensemble $\mathfrak{C}(d_c, d_v)$ with check degree $d_c$ and bit degree $d_v$ can only achieve those rate-distortion pairs $(R, D)$ that satisfy the bound*

$$R \left[ 1 - \max_{\delta \in (0,1)} \min \{g(\delta, \beta),\ (1-\delta)\beta\} \right] \geq 1 - H(D), \qquad (6)$$

*where $\beta := \frac{d_v!}{2} \left( \frac{D}{d_v} \right)^{d_v}$, and $g(\delta, \beta) := \log_2(e) \left[ \delta^2 \frac{\beta^2}{2 d_v^2 R(1-D)} \right]$. For any finite bit degree $d_v$, the minimal rate $R$ satisfying the relation (6) is strictly bounded away from the Shannon rate-distortion bound $R(D) = 1 - H(D)$.*

Figure 2 (b) again illustrates[1] the excess rate guaranteed by Corollary 2 for different distortions, this time as a function of the bit degrees $d_v$ in this construction. At least on the basis of the tight-ness of these bounds, the bit-check-regular ensemble appears to have rate-distortion performance superior to the check-regular ensemble, which is to be expected intuitively.

---

[1]Strictly speaking, the plots in Figure 2(b) is misleading, in that not all rates shown can be achieved for every degree. For instance, for degree $d_v = 3$, it is only possible to achieve rate 2/3 with a regular degree distribution. However, we gloss over this technical issue for the sake of clearer comparison with Figure 2(a).

Note that in both Corollaries 1 and 2, the bounds are guaranteed to hold with high probability over a choice of random code from the ensemble. Therefore, although these results guarantee that almost all codes from the given ensembles are bounded away from the Shannon limit, they do not rule out the possibility that there exists some fixed code of the specified type that achieves the Shannon limit. In the initial conference version of this paper [6], we conjectured that no such codes exist—i.e., that every code with degrees chosen according the specified ensemble must satisfy the given bounds. In recent work, Kudekar and Urbanke [12] used related methods to establish lower bounds that hold for *fixed* bounded degree codes, as opposed to the random ensemble results given here.

## 3 Tools for lower bounding rate-distortion

In this section, we develop the analytical tools that underlie the proof of Theorem 1, as a prelude to our analysis of random ensembles to follow in Section 4. We begin by describing a certain type of $D$-ball encoder, in general sub-optimal relative to the optimal encoder, but more amenable to analysis. As a first step, we establish the asymptotic optimality of this $D$-ball encoding. We then exploit this encoder to prove Theorem 1.

### 3.1 Maximum likelihood and $D$-ball encoding

Recall that any binary linear code $\mathbb{C} \subset \{0,1\}^n$ can be characterized by a generator matrix $\mathbf{G} \in \{0,1\}^{n \times m}$, such that any codeword $x \in \mathbb{C}$ is of the form $x = \mathbf{G}z$, where $z \in \{0,1\}^m$ is a sequence of information bits. Given some source sequence $S \in \{0,1\}^n$ of symmetric Bernoulli random variables, suppose that we quantize the source using the code $\mathbb{C}$. We use $\widehat{Z} \in \{0,1\}^m$ to denote the random information sequence, with associated codeword $\widehat{S} = \mathbf{G}\widehat{Z}$, to which the random sequence $S$ is quantized.

The optimal encoding map, from source sequences $S$ to codewords $\widehat{Z}$, is the so-called maximum likelihood (ML) encoder. Given a source sequence, it computes the set of information sequences that generate codewords closest to $S$ in Hamming distance, and outputs one of them uniformly at random—say $\widehat{Z}_{\mathrm{ML}} \in \arg\min_{z \in \{0,1\}^m} \{\|\mathbf{G}z \oplus S\|_1\}$. The associated minimal distortion is a random variable, defined as

$$d_n(S; \mathbb{C}) \quad := \quad = \frac{1}{n} \min_{z \in \{0,1\}^m} \{\|\mathbf{G}z \oplus S\|_1\} \;=\; \frac{1}{n} \|\mathbf{G}\widehat{Z}_{\mathrm{ML}} \oplus S\|_1. \tag{7}$$

This ML encoder is optimal in that its expected distortion $\mathbb{E}[d_n(S; \mathbb{C})]$ is minimized over all encoders.

Despite the optimality of ML encoding, it is more convenient for theoretical purposes to analyze the following $D$-ball encoder. For any fixed target distortion $D \in (0, \frac{1}{2})$, define the Hamming ball of radius $D$ around the source sequence $S$ as follows:

$$\mathbb{B}_n(S; D) := \{x \in \{0,1\}^n \mid \|S \oplus x\|_1 \leq Dn\}. \tag{8}$$

We say that the $D$-ball encoder *succeeds* if and only if the intersection $\mathbb{B}_n(S; D) \cap \mathbb{C}$ is non-empty, in which case it chooses some information sequence $\widehat{Z}_{\text{DB}}$ uniformly at random from the set $\{z \in \{0,1\}^m \mid \mathbf{G}z \in \mathbb{B}_n(S; D)\}$. Otherwise, the encoder fails, and we set $\widehat{Z}_{\text{DB}} = z^*$, where $z^*$ is some arbitrary non-zero sequence. We claim that this $D$-ball encoder is asymptotically equivalent to the ML encoder.

**Lemma 1.** *For any binary linear code, the following two conditions are equivalent:*

(a) *for all $\epsilon > 0$, the probability of success under $(D + \epsilon)$-ball encoding converges to one as $n \to +\infty$.*

(b) *for all $\delta > 0$, we have $\mathbb{E}[d_n(S; \mathbb{C})] \leq D + \delta$ for all suitably large blocklengths $n$.*

*Proof:* We first show that (a) implies (b). Given any fixed $\delta > 0$, set $\epsilon = \delta/2$ in part (a), and consider the associated $(D + \frac{\delta}{2})$-ball encoder. Setting $p_n = \mathbb{P}[(D + \delta/2)\text{-ball success}]$ and $\widehat{S}_{\text{DB}} = \mathbf{G}\widehat{Z}_{\text{DB}}$, we have

$$\begin{aligned}
\frac{1}{n}\mathbb{E}[\|\widehat{S}_{\text{DB}} \oplus S\|_1] &\leq (D + \frac{\delta}{2})p_n + (1 - p_n)\frac{1}{2} \\
&\leq D + \frac{1}{2}[1 - p_n + p_n\delta] \\
&\leq D + \delta,
\end{aligned}$$

where the final inequality follows if we can ensure that $p_n \geq \frac{1-2\delta}{1-\delta}$. Since $p_n \to 1$ by assumption, this condition can be met by choosing $n$ sufficiently large. Finally, since ML encoding yields the minimal average distortion, we have $\mathbb{E}[d_n(S;\mathbb{C})] \leq \frac{1}{n}\mathbb{E}[\|\widehat{S}_{\text{DB}} \oplus S\|_1] \leq D + \delta$, which is the claim (b).

We now prove that {not (a)} implies {not (b)}. Suppose that for some $\epsilon > 0$, the encoding success probability $p_n = \mathbb{P}[(D + \epsilon)\text{-ball success}]$ does not converge to 1. Then $\liminf p_n < 1$, so that by taking subsequences if necessary, we may assume that for all sufficiently large $n$, the failure probability satisfies $1 - p_n \geq \nu$ for some $\nu > 0$. Since the $(D+\epsilon)$-ball encoder can fail only if there are no codewords within normalized distance $(D+\epsilon)$ of the source sequence, this statement implies $\mathbb{P}[d_n(S;\mathbb{C}) > D + \epsilon] \geq \nu$.

Next, we claim that the ML distortion $d_n(S;\mathbb{C})$ is concentrated around its expected value. Consider the martingale sequence based on exposing the source bits in the order $S_1, S_2, \ldots, S_n$, and defining the sequence of random variables $Z_0 = \mathbb{E}[d_n(S; \mathbb{C})]$, and

$$Z_k := \mathbb{E}[d_n(S; \mathbb{C}) \mid S_1, \ldots S_k], \qquad \text{for } k = 1, 2, \ldots, n,$$

such that $Z_n = d_n(S; \mathbb{C})$. Note that changing one bit $S_i$ changes the (normalized) distortion $d_n(S; \mathbb{C})$ by at most $1/n$, which means that $d_n(S; \mathbb{C})$ is a $c$-Lipschitz function with constant $c = 1/n$. Therefore, by applying the Azuma-Hoeffding inequality [10] to this martingale sequence yields that $\mathbb{P}\left[|d_n(S; \mathbb{C}) - \mathbb{E}[d_n(S; \mathbb{C})]| \geq \epsilon\right] \leq 2\exp\left(-\frac{n\epsilon^2}{2}\right)$. Therefore, for any constant $\epsilon > 0$, we have

$$\lim_{n \to +\infty} \mathbb{P}\left[|d_n(S; \mathbb{C}) - \mathbb{E}[d_n(S; \mathbb{C})]| \geq \epsilon\right] = 0. \tag{9}$$

Using the sharp concentration (9), we see that the bound $\mathbb{P}[d_n(S; \mathbb{C}) > D + \epsilon] \geq \nu$ implies that $\mathbb{E}[d_n(S; \mathbb{C})] \geq D + \epsilon/2$ w.h.p. Hence, we have established the existence of some $\epsilon > 0$ for which there exists an infinite sequence of blocklengths $n$ along which $\mathbb{E}[d_n(S; \mathbb{C})] \geq D + \epsilon/2$, thus implying {not (b)}. ∎

### 3.2 Proof of Theorem 1

We are now equipped to prove Theorem 1. If the code $\mathbb{C}$ achieves average distortion $D$ by some encoding method, then Lemma 1 implies that the $D$-ball encoder must achieve this distortion. Letting $A(S, D; \mathbb{C})$ be the event that the $D$-ball encoder succeeds for source sequence $S$, denote

$$p_n = \mathbb{P}[A(S, D; \mathbb{C})].$$

Recall the operation of the $D$-ball encoder: when it succeeds—that is, for any source sequence $S$ such that $N(S, D; \mathbb{C}) \geq 1$—the encoder chooses an information sequence $\widehat{Z}$ uniformly at random from all information sequences satisfying $\mathbf{G}\widehat{Z} = \widehat{S}$. Note that there are $N(S, D; \mathbb{C})$ such choices, by the definition (3) of $N$. Otherwise, if $D$-ball encoding fails, the encoder simply chooses some fixed non-zero information sequence $z^* \neq \vec{0}$.

By definition of this decoder, for any source sequence $s$ for which the $D$-ball encoding succeeds, we have

$$\mathbb{P}[\widehat{Z} = \widehat{z} \mid S = s] = \begin{cases} \frac{1}{N(s, D; \mathbb{C})} & \text{if } \mathbf{G}\widehat{z} \in \mathbb{B}_n(s; D) \\ 0 & \text{otherwise.} \end{cases} \tag{10}$$

We now compute this PMF of the random variable $\widehat{Z}$.

**Lemma 2.** *The PMF of $\widehat{Z}$ has the form*

$$\mathbb{P}[\widehat{Z} = z] = \begin{cases} q(D;\mathbb{C}) & \text{if } z \neq z^*, \text{ and} \\ q(D;\mathbb{C}) + (1-p_n) & \text{for } z = z^*. \end{cases} \quad (11)$$

*where* $q(D;\mathbb{C}) := \sum_{u \in \mathbb{B}_n(\vec{0};D)} \frac{1}{N(u,D;\mathbb{C})} 2^{-n}$ *and* $p_n = \mathbb{P}[A(S,D;\mathbb{C})]$.

*Proof:* For any $z \in \{0,1\}^m$ (not equal to the special sequence $z^*$), we have

$$\mathbb{P}[\widehat{Z}=z] = \sum_s \mathbb{P}[\widehat{Z}=z \mid S=s]\mathbb{P}[S=s] = \sum_{\{s \mid \|\mathbf{G}z \oplus s\|_1 \leq Dn\}} \frac{1}{N(s,D;\mathbb{C})} 2^{-n},$$

using the form of the PMF (10), and the fact $\mathbb{P}[S=s] = 2^{-n}$ for all source sequences.

Let $t \neq z^*$ be any other information sequence. Then

$$\mathbb{P}[\widehat{Z}=z] = \sum_{\{s \mid \|\mathbf{G}z \oplus s\|_1 \leq Dn\}} \frac{1}{N(s,D;\mathbb{C})} 2^{-n}$$

$$= \sum_{\{s \mid \|\mathbf{G}t \oplus (s \oplus \mathbf{G}(z \oplus t))\|_1 \leq Dn\}} \frac{1}{N(s,D;\mathbb{C})} 2^{-n}$$

$$= \sum_{\{s' \mid \|\mathbf{G}t \oplus s'\|_1 \leq Dn\}} \frac{1}{N(s' \oplus \mathbf{G}(z \oplus t), D;\mathbb{C})} 2^{-n},$$

where we have defined $s' := s \oplus \mathbf{G}(z \oplus t)$.

We now claim that from the symmetry of the code, for any codeword $x_0 \in \mathbb{C}$ and source sequence $s \in \{0,1\}^n$, we have $N(s,D;\mathbb{C}) = N(s \oplus x_0, D;\mathbb{C})$. Indeed, suppose that $N(s,D;\mathbb{C}) = k$, with codewords $x_1, \ldots, x_k$ such that $\|x_i \oplus s\|_1 \leq Dn$ for $i = 1, \ldots, k$. Then the codewords $x_i' := x_i \oplus x_0$ satisfy

$$\|x_i' \oplus (s \oplus x_0)\|_1 = \|(x_i \oplus x_0) \oplus (s \oplus x_0)\|_1 = \|x_i \oplus s\|_1 \leq Dn,$$

so that $N(s \oplus x_0, D;\mathbb{C}) \geq k$, and by symmetry $N(s \oplus x_0, D;\mathbb{C}) = k$.

Consequently, we have $N(s' \oplus \mathbf{G}(z \oplus t), D;\mathbb{C}) = N(s',D;\mathbb{C})$, so that for $z \neq z^*$, we have

$$\mathbb{P}[\widehat{Z}=z] = \sum_{\{s' \mid \|\mathbf{G}t \oplus s'\|_1 \leq Dn\}} \frac{1}{N(s' \oplus \mathbf{G}(z \oplus t), D;\mathbb{C})} 2^{-n}$$

$$= \sum_{\{s' \mid \|\mathbf{G}t \oplus s'\|_1 \leq Dn\}} \frac{1}{N(s',D;\mathbb{C})} 2^{-n}.$$

This statement holds for any $t \neq z^*$, so that setting $t = 0$ yields the first part of the claim (11).

Finally, for $z = z^*$, we have

$$\mathbb{P}[\widehat{Z} = z^*] = q(D;\mathbb{C}) + \sum_{\{s | A(s, D; \mathbb{C}) \text{ not true}\}} 2^{-n} = q(D;\mathbb{C}) + (1 - p_n),$$

which completes the proof. ∎

We have expressed the probability of selecting a compressed sequence $z$ (and hence codeword), as a function of the overlaps $q(D;\mathbb{C})$ between the $D$-balls of codewords. The key point now is that if there are large overlaps (i.e. if $N$ is large for many sequences), then more codewords will be needed to cover the total number $2^n$ of possible binary sequences, and hence there will be a rate loss. This intuitive argument can be made formal by observing that the PMF of $\widehat{Z}$ needs to be normalized—namely, we must have $\sum \mathbb{P}[\widehat{Z} = z] = 1$. Using the PMF of $\widehat{Z}$ from Lemma 2, we obtain $2^{nR} q(D;\mathbb{C}) + (1 - p_n) = 1$, or equivalently (upon solving for $p_n$):

$$p_n = 2^{nR} \sum_{u \in \mathbb{B}_n(\vec{0};D)} \frac{1}{N(u, D; \mathbb{C})} 2^{-n}.$$

Taking logarithms yields

$$\frac{1}{n} \log p_n = R - 1 + \frac{1}{n} \log \sum_{u \in \mathbb{B}_n(\vec{0};D)} \frac{1}{N(u, D; \mathbb{C})}$$

$$= R - 1 - \frac{1}{n} \log \binom{n}{Dn} + \frac{1}{n} \log \frac{1}{\binom{n}{Dn}} \sum_{u \in \mathbb{B}_n(\vec{0};D)} \frac{1}{N(u, D; \mathbb{C})},$$

where the last equality holds by adding and subtracting $\log \binom{n}{Dn}$, corresponding to (the logarithm of) the total number of binary sequences $|\{\mathbb{B}_n(\vec{0}; D)\}|$ within the $D$-ball centered at zero.

By construction, the last term is just an expectation over a uniformly selected sequence $U$ in $\mathbb{B}_n(\vec{0}; D)$, as defined prior to the statement of Theorem 1, so that we have

$$\frac{1}{n} \log p_n = R - 1 + \frac{1}{n} \log \binom{n}{Dn} + \frac{1}{n} \log \mathbb{E}_U[1/N(U, D; \mathbb{C})]. \tag{12}$$

This expression is the exact exponent of the success probability of the $D$-ball decoder and might be of independent interest. If this exponent is negative, the probability of success of the $D$-ball encoder will vanish exponentially.

Our final step is to apply standard asymptotics for binomial coefficients [4]—namely, $\frac{\log \sum_{k=0}^{Dn} \binom{n}{k}}{n} = H(D) \pm o(1)$. Substituting into equation (12), we obtain that the probability of $D$-ball success van-ishes exponentially quickly if

$$R < 1 - \frac{1}{n}\log\left|\{\mathbb{B}_n(\vec{0};D)\}\right| - \frac{1}{n}\log \mathbb{E}_U[1/N(U,D;\mathbb{C})]. \tag{13}$$

By Lemma 1, since the D-ball encoder fails for this rate-distortion pair, the ML encoder must also fail to achieve average distortion D, which establishes Theorem 1. The last term, that describes the expected overlaps of $D$-balls, is the only term that depends on the specific code used, and corresponds to the excess rate due to the code suboptimality.

## 4 Analysis over random ensembles

From the lower bound (4), we see that any loss relative to the Shannon limit is captured by the *excess rate coefficient*—namely, $\log \mathbb{E}_U[1/N(U,D;\mathbb{C})]/n$. For a fixed code $\mathbb{C}$, a challenge associated with analysis of this quantity is the possible non-uniformity in the cardinalities

$$N(u,D;\mathbb{C}) = \left|\{z \in \{0,1\}^m \mid \mathbf{G}z \in \mathbb{B}_n(u;D)\}\right|.$$

As a concrete example, consider the rate $R = 1/2$ code with $(m,n) = (2,4)$ and codewords

$$\mathbb{C} = \left\{\begin{bmatrix} 0 & 0 & 0 & 0 \end{bmatrix}, \begin{bmatrix} 1 & 1 & 0 & 0 \end{bmatrix}, \begin{bmatrix} 0 & 0 & 0 & 1 \end{bmatrix}, \begin{bmatrix} 1 & 1 & 0 & 1 \end{bmatrix}\right\}.$$

With $Dn = 1$, a simple calculation shows that for this code,

$$N(u,D;\mathbb{C}) = \begin{cases} 1 & \text{for } u = \begin{bmatrix} 0 & 0 & 1 & 0 \end{bmatrix} \\ 2 & \text{otherwise.} \end{cases}$$

Consequently, the quantity $1/N(u,D;\mathbb{C})$ is directionally biased towards the quantization noise sequence $u = \begin{bmatrix} 0 & 0 & 1 & 0 \end{bmatrix}$.

Although evaluating the excess rate coefficient for a fixed code appears difficult, if instead we view $\mathbb{C}$ as a random variable, drawn from some ensemble $\mathfrak{C}$ of codes, then the excess rate becomes a random variable, as a function of the random code $\mathbb{C}$. We can then consider ensemble-based analysis of this random variable.

### 4.1 Concentration and graph-based certificate

We begin by stating conditions involving expectations and concentration that are sufficient to yield bounds (holding with high probability) on the rate-distortion of a randomly drawn code $\mathbb{C}$. In this analysis, we consider random ensembles $\mathfrak{C}$, in which the code bits are exchangeable, meaning that the probability distribution is invariant to permutations of the labelings of the code bits. For instance, this exchangeability holds for the check-regular and bit-check-regular ensembles defined in Section 2.2.

**Proposition 1.** *Given an exchangeable ensemble $\mathfrak{C}$, define the random variable*

$$W(D; \mathbb{C}) := \big|\{z \in \{0,1\}^m \mid (\mathbf{G}z)_i = 0 \text{ for all } i \notin \{1, 2, \ldots, Dn\}\}\big|. \tag{14}$$

*and suppose that $\frac{1}{n}\mathbb{E}[\log W(D; \mathbb{C})] \geq \alpha(\mathfrak{C}) > 0$, and moreover that for all $\delta \in (0, 1)$,*

$$\mathbb{P}[\log W(D; \mathbb{C}) \leq (1 - \delta)\alpha(\mathfrak{C})n] \leq K\, 2^{-f(\delta)n}, \tag{15}$$

*for some positive constant $K \equiv K(\mathfrak{C})$ independent of blocklength, and positive function $f : [0, 1] \to (0, \infty)$. Then with probability converging to one as $n \to +\infty$, a randomly drawn code $\mathbb{C}$ cannot satisfy any rate-distortion pair $(R, D)$ for which*

$$R < 1 - H(D) + \max_{\delta \in (0,1)} \min\{f(\delta),\, (1-\delta)\alpha(\mathfrak{C})\} - o(1). \tag{16}$$

*Proof:* For each $u \in \mathbb{B}_n(\vec{0}; D)$, define the random variable

$$M(u, D; \mathbb{C}) := \big|\{z \in \{0,1\}^m \mid \mathbf{G}z \in \mathbb{B}_n(\vec{0}; D) \cap \mathbb{B}_n(u; D)\}\big|, \tag{17}$$

and note that $\frac{1}{N(u,D;\mathbb{C})} \leq \frac{1}{M(u,D;\mathbb{C})}$ by construction. Using this fact, and applying Jensen's inequality with the concavity of the logarithm, we have

$$\mathbb{E}_{\mathbb{C}}\left[\frac{\log \mathbb{E}_U[1/N(U, D; \mathbb{C})]}{n}\right] \leq \mathbb{E}_{\mathbb{C}}\left[\frac{\log \mathbb{E}_U[1/M(U, D; \mathbb{C})]}{n}\right]$$

$$\leq \frac{\log \mathbb{E}_{\mathbb{C}}\mathbb{E}_U[1/M(U, D; \mathbb{C})]}{n}.$$

For an exchangeable ensemble of codes, the distribution of $M(u, D; \mathbb{C})$ depends only on Hamming weight $\|u\|_1$, so that we can write

$$\frac{\log \mathbb{E}_{\mathbb{C}}\mathbb{E}_U[1/M(U, D; \mathbb{C})]}{n} = \frac{1}{n} \log \frac{\sum_{k=0}^{Dn} \binom{n}{k} \mathbb{E}_{\mathbb{C}}[1/M(u^k, D; \mathbb{C})]}{\binom{n}{Dn}}$$

where for each $k = 0, 1, \ldots, Dn$, the vector $u^k$ has Hamming weight $\|u^k\|_1 = k$, with $u^0 = \vec{0}$ and for $k \geq 1$,

$$(u^k)_i := \begin{cases} 1 & \text{for } i = 1, \ldots, k \\ 0 & \text{otherwise.} \end{cases}$$

Now observe that by the definition (14) of $W$ and $u^k$, we have $W(D; \mathbb{C}) \leq M(u^k, D; \mathbb{C})$ for all

$k = 0, 1, \ldots, Dn$, whence

$$\mathbb{E}_\mathbb{C} \frac{\log \mathbb{E}_U[1/N(U, D; \mathbb{C})]}{n} \leq \frac{\log \mathbb{E}_\mathbb{C} \mathbb{E}_U[1/M(U, D; \mathbb{C})]}{n} \leq \frac{1}{n} \log \mathbb{E}_\mathbb{C}[1/W(D; \mathbb{C})].$$

Combining this bound with Theorem 1, we have shown that the ensemble rate-distortion must satisfy

$$R \geq 1 - H(D) - \frac{1}{n} \log \mathbb{E}_\mathbb{C}[1/W(D; \mathbb{C})] - o(1).$$

To conclude the proof, we now exploit the given assumptions on the behavior of $\log W(D; \mathbb{C})$. Defining the event $A(\delta) := \{1/W(D; \mathbb{C}) \geq 2^{-n(1-\delta)\alpha(\mathfrak{C})}\}$, we have $\mathbb{P}[A(\delta)] \leq K 2^{-nf(\delta)}$ from the concentration (15). Since $1/W(D; \mathbb{C}) \leq 1$, we can write

$$\frac{1}{n} \log \mathbb{E}[1/W(D; \mathbb{C})] \leq \frac{1}{n} \log \left\{ K 2^{-nf(\delta)} + 2^{-n(1-\delta)\alpha(\mathfrak{C})} \right\}$$
$$\leq -\min\{f(\delta), (1-\delta)\alpha(\mathfrak{C})\} + o(1),$$

since with $K$ independent of blocklength, we have $\log K/n = o(1)$. ∎

Proposition 1 suggests a general procedure for proving lower bounds on the effective rate-distortion of different random ensembles of codes, by controlling the behavior of the random variable $W(D; \mathbb{C})$. In order to do so, it is convenient to make use of the following graph-based certificate. Given the factor graph describing the generator matrix $\mathbf{G}$ of the code $\mathbb{C}$, suppose that the last $(1 - D)n$ checks are labeled as fixed, denoted by $\mathbb{S}^\text{fix}$. We use $N(\mathbb{S}^\text{fix}; \mathbb{C})$ to denote the subset of information bits connected to at least one check in $\mathbb{S}^\text{fix}$, and let $\mathbb{T}^\text{free}(\mathbb{C})$ denote the complement of these fixed information bits. See Figure 3 for an illustration of these concepts.

The key property of this construction is the following: suppose that we set $z_i = 0$ for all indices $i \in N(\mathbb{S}^\text{fix}; \mathbb{C})$. With this setting, the information bits $z_j$ associated with indices $j \in \mathbb{T}^\text{free}(\mathbb{C})$ can be altered arbitrarily, while still ensuring that the codeword $\mathbf{G}z$ still satisfies $(\mathbf{G}z)_\ell = 0$ for all $\ell \in \mathbb{S}^\text{fix}$. The *number of free bits* $F(\mathbb{C}) := |\{\mathbb{T}^\text{free}(\mathbb{C})\}|$ thereby provides a lower bound on $\log W(D; \mathbb{C})$, one which is relatively easy to analyze for random ensembles. This graph-based certificate suggests the following two-stage approach for generating lower bounds:

(a) First establish a lower bound on $\mathbb{E}[F(\mathbb{C})]$, and thus a lower bound on $\mathbb{E}[\log W(D; \mathbb{C})]$.

(b) Next show that $F(\mathbb{C})$ and consequently $\log W(D; \mathbb{C})$ is larger than this lower bound with high probability.

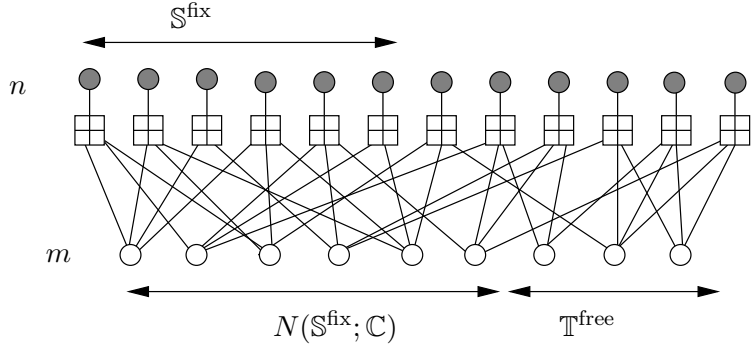

**Figure 3.** Factor graph of an LDGM code illustrating the fixed checks $\mathbb{S}^{\text{fix}}$, information bit neighbors $N(\mathbb{S}^{\text{fix}}; \mathbb{C})$ of fixed checks, and the free information bits $\mathbb{T}^{\text{free}}$.

### 4.2 Lower bounds for specific ensembles

We now illustrate this approach by using it to prove Corollary 1 for the check-regular ensemble $\mathfrak{C}(d_c)$, and then to prove Corollary 2 for the bit-check-regular ensemble $\mathfrak{C}(d_c, d_v)$. (See Section 2.2 for definitions of these ensembles.) With appropriate modifications, the underlying ideas of our approach should be more generally applicable, for instance to LDGM codes with irregular degree distributions.

#### 4.2.1 Proof of Corollary 1

Recall the check-regular ensemble of LDGM codes, denoted by $\mathfrak{C}(d_c)$: it consists of codes $\mathbb{C}(d_c)$ with $n$ checks and $m$ information bits, constructed by having each check select $d_c$ bits uniformly at random (and with repetition). We have the following result concerning the expected number of free bits:

**Lemma 3.** *The expected number of free bits over the check-regular ensemble grows linearly*

$$\frac{\mathbb{E}[F(\mathbb{C})]}{m} = \left(1 - \frac{1}{m}\right)^{(1-D)nd_c} =: \beta \qquad (18)$$

*and moreover, $F(\mathbb{C})$ is sharply concentrated, in that for all $\delta \in (0,1)$, we have*

$$\mathbb{P}\left[F(\mathbb{C}) \leq (1-\delta)\mathbb{E}[F(\mathbb{C})]\right] \leq 2\exp\left\{-m\left[(1-\beta) + \delta\beta\right]\log\left[1 + \frac{\delta\beta}{1-\beta}\right]\right\}. \qquad (19)$$

*Proof:* Any particular information bit is adjacent to a particular check with probability $(\frac{m-1}{m})^{d_c}$, and this event occurs for each of the $(1-D)n$ fixed checks independently. The probability that a particular bit is free (i.e. non-adjacent to $\mathbb{S}^{\text{fix}}$) is simply $\beta$ as defined in equation (18), and the the expected size of $\mathbb{T}^{\text{free}}$ is simply $\mathbb{E}[F(\mathbb{C})] = m\beta$ as claimed. Now we need to show that the random variable $F(\mathbb{C})$ is concentrated around its mean. Since $F(\mathbb{C})$ is a sum of $m$ i.i.d. Bernoulli variables with mean $\beta$, Sanov's theorem [4] yields that for any $\delta \in (0,1)$,

$$\mathbb{P}\left[F(\mathbb{C}) \leq (1-\delta)\beta\right] \leq 2 \exp\left\{-m \text{ KL}((1-\delta)\beta \| \beta)\right\},$$

where $\text{KL}(a\|b)$ is the Kullback-Leibler divergence for Bernoulli variates. Noting the lower bound

$$\text{KL}((1-\delta)\beta \| \beta) \geq [(1-\beta) + \delta\beta] \log\left\{1 + \frac{\delta\beta}{1-\beta}\right\},$$

claim (19) follows. ∎

Using Lemma 3 and Proposition 1, we can now prove Corollary 1. Assume that for some pair $(R, D)$ and code $\mathbb{C}$ drawn from the check-regular ensemble, the source encoder is successful. Recall that $F(\mathbb{C})$ is a lower bound on $\log W(D; \mathbb{C})$. Using the fact that $m = Rn$, equation (18) implies that

$$\frac{1}{n}\mathbb{E}[\log W(D;\mathbb{C})] \geq R\left(1 - \frac{1}{m}\right)^{(1-D)nd_c} := \alpha(\mathfrak{C}) = R\beta.$$

Moreover, we have

$$\begin{aligned}
\mathbb{P}[\log W(D;\mathbb{C}) \leq (1-\delta)\alpha(\mathbb{C})n] &\leq \mathbb{P}[F(\mathbb{C}) \leq (1-\delta)\beta m] \\
&= \mathbb{P}[F(\mathbb{C}) \leq (1-\delta)\mathbb{E}[F(\mathbb{C})]] \\
&\leq 2 \exp\left\{-m[(1-\beta) + \delta\beta] \log\left[1 + \frac{\delta\beta}{1-\beta}\right]\right\}
\end{aligned}$$

using equation (19). Consequently, the hypotheses of Proposition 1 are satisfied with $K = 2$, $\beta = \left(1 - \frac{1}{m}\right)^{(1-D)nd_c}$, $\alpha(\mathfrak{C}) = R\beta$, and

$$f(\delta; \beta) := \log_2(e) R [(1-\beta) + \delta\beta] \log\left[1 + \frac{\delta\beta}{1-\beta}\right] := R g(\delta; \beta),$$

so that Proposition 1 implies that

$$\begin{aligned}
R &\geq 1 - H(D) + R \max_{\delta \in (0,1)} \min\left\{g(\delta; \beta), (1-\delta)\beta\right\} - o(1) \\
&\geq 1 - H(D) + \frac{1}{2}R\beta - o(1),
\end{aligned} \quad (20)$$

where it can be verified numerically that for $\delta = 0.5$, we have $g(\delta; \beta) \geq (1-\delta)\beta$ for all $\beta \in [0,1]$. Finally, a standard Poisson limit yields

$$\lim_{n \to +\infty} \left(1 - \frac{1}{Rn}\right)^{(1-D)nd_c} = \exp\left(-\frac{(1-D)d_c}{R}\right),$$

so that $\beta \geq \exp\left(-\frac{(1-D)d_c}{R}\right) - o(1)$. Therefore, for all $n$ sufficiently large, the pair $(R, D)$ must satisfy

$$R\left[1 - \frac{1}{2}\exp\left(-\frac{(1-D)d_c}{R}\right)\right] \geq 1 - H(D),$$

as claimed.

### 4.2.2 Proof of Corollary 2

We now turn of the proof of Corollary 2, concerning the effective rate-distortion function of the bit-check-regular ensemble $\mathfrak{C}(d_c, d_v)$ of codes. We begin by addressing the expected value and concentration of the random variable $F(\mathbb{C})$ for this ensemble.

**Lemma 4.** *The expectation of $F(\mathbb{C})$ grows linearly in blocklength: in particular, it is lower bounded as*

$$\frac{\mathbb{E}[F(\mathbb{C})]}{m} \geq \frac{d_v!}{2}\left(\frac{D}{d_v}\right)^{d_v} := \beta. \tag{21}$$

*Moreover, it is sharply concentrated in that for all $\delta \in (0, 1)$,*

$$\mathbb{P}\left[F(\mathbb{C}) \leq (1-\delta)\mathbb{E}[F(\mathbb{C})]\right] \leq 2\exp\left\{-m\delta^2 \frac{\beta^2}{2d_v^2 R(1-D)}\right\}. \tag{22}$$

*Proof:* Any code in the $\mathfrak{C}(d_c, d_v)$ ensemble is characterized by a set of $d_c n = d_v m$ edges, matching the $n$ checks to the $m$ information bits. A random code is generated by selecting a permutation $\pi$ of the $d_c n$ edges, uniformly at random from all $(d_c n)!$ such permutations. For a fixed information bit, let $N_{\text{good}}$ denote the number of permutations where all the $d_v$ edges of a particular information bit are adjacent to non-fixed checks. Since there are $n(1-D)$ fixed checks and $Dn$ free checks, there are $n(1-D)d_c$ edges adjacent to fixed checks. Consequently, the probability that a particular information bit $i \in \{1, \ldots, m\}$ is free is simply $q = N_{\text{good}}/(d_c n)!$.

To determine $q$ more explicitly, we count the number $N_{\text{good}}$ of permutations that leave a particular information bit $i$ free. Since there are $Dd_c n$ (labeled) edges adjacent to free checks, there are $\binom{Dd_c n}{d_v}d_v!$ ways for the given information bit to connect all of its $d_v$ outgoing edges to such checks. The remaining $nd_c - d_v$ edges can be permuted arbitrarily without affecting the connectivity of the given information bit, which produces another factor of $(nd_c - d_v)!$. Overall, we conclude that

$$q = \frac{d_v!\binom{Dd_c n}{d_v}(nd_c - d_v)!}{(nd_c)!}. \tag{23}$$

In order to show that the expectation $\mathbb{E}[F(\mathbb{C})]$ scales linearly, it suffices to show that $q$ is lower bounded by a constant. We use the following bounds [20] that follow from Stirling's approximation:

$$\sqrt{2\pi m}\left(\frac{m}{e}\right)^m \leq m! \leq 2\sqrt{2\pi m}\left(\frac{m}{e}\right)^m. \tag{24}$$

We also require a lower bound on the binomial coefficient $\binom{n}{k}$; one such bound is

$$\binom{n}{k} \geq \left(\frac{n}{k}\right)^k = C(k)n^k \tag{25}$$

where $C(k) = (1/k)^k$ for all positive integers $k$. Using these bounds, we obtain

$$(nd_c)! \leq 2\sqrt{2\pi nd_c}\left(\frac{nd_c}{e}\right)^{nd_c}, \quad \text{and}$$

$$(nd_c - d_v)! \geq \sqrt{2\pi(nd_c - d_v)}\left(\frac{nd_c - d_v}{e}\right)^{nd_c - d_v}.$$

Consequently, we have

$$q \geq \frac{(d_v)! C(d_v)(Dd_c)^{d_v} n^{d_v}\left(\frac{nd_c - d_v}{e}\right)^{nd_c - d_v}}{\left(\frac{nd_c}{e}\right)^{nd_c}} \cdot \frac{\sqrt{2\pi(nd_c - d_v)}}{2\sqrt{2\pi(nd_c)}}.$$

We have $\frac{\sqrt{2\pi(nd_c - d_v)}}{2\sqrt{2\pi(nd_c)}} = \frac{1}{2} - o(1)$ as $n \to +\infty$ with $d_v$ and $d_c$ fixed, so that after some further algebra, we obtain

$$q \geq \left\{\frac{1}{2} - o(1)\right\}\left(C(d_v)\,(d_v)!(Dd_c)^{d_v}\right)\left(\frac{e}{d_c}\right)^{d_v}\left(1 - \frac{d_v}{d_c n}\right)^{nd_c}$$

$$= \left\{\frac{1}{2} - o(1)\right\}\left(C(d_v)\,(d_v)!\,D^{d_v}\right)\exp(d_v)\left(1 - \frac{d_v}{d_c n}\right)^{nd_c}$$

$$= \left\{\frac{1}{2} - o(1)\right\}\left(C(d_v)\,(d_v)!\,D^{d_v}\right)\{1 - o(1)\},$$

where the final line follows since $\left(1 - \frac{d_v}{d_c n}\right)^{nd_c}$ converges to $e^{-d_v}$ as $n$ tends to infinity. Overall, since the quantity $C(d_v)\,(d_v)!\,D^{d_v} = (d_v)!\left(\frac{D}{d_v}\right)^{d_v}$ stays bounded from above for all $d_v \geq 2$, we obtain the final lower bound

$$q \geq \frac{(d_v)!}{2}\left(\frac{D}{d_v}\right)^{d_v} - o(1) \tag{26}$$

as $n \to +\infty$. Since there are $m$ information bits in total, this bound with linearity of expectation establishes the lower bound (21) on the expected value.

Finally, we establish the concentration (22) of $F(\mathbb{C})$ for this ensemble. By definition, the random variable $F(\mathbb{C})$ is completely specified by the edge sets of fixed checks indexed by the set $\mathbb{S}^{\text{fix}}$ of cardinality $(1-D)n$, as in Figure 3. For $i = 1, \ldots, (1-D)n$, let $V_i$ be a random variable specifying the edge set of fixed check $i \in \mathbb{S}^{\text{fix}}$, and define the martingale sequence $Z_0 = \mathbb{E}[F(\mathbb{C})]$, and $Z_i = \mathbb{E}[F(\mathbb{C}) \mid V_1, \ldots V_i]$, such that $V_{(1-D)n} = F(\mathbb{C})$. Since each check has degree $d_c$, we have the bound $|Z_{i+1} - Z_i| \leq d_c$. Therefore, by the Azuma-Hoeffding inequality [20], we have

$$\mathbb{P}[|F(\mathbb{C}) - \mathbb{E}[F(\mathbb{C})]| \geq \epsilon(1-D)n] \leq 2 \exp\left\{-n\epsilon^2(1-D)/(2d_c^2)\right\}.$$

Setting $\epsilon = \delta \frac{\beta m}{(1-D)n} = \delta \frac{\beta R}{(1-D)}$ yields that

$$\mathbb{P}[F(\mathbb{C}) \leq (1-\delta)\mathbb{E}[F(\mathbb{C})]] \leq 2 \exp\left\{-n\delta^2 \frac{\beta^2 R^2}{(1-D)^2} \frac{(1-D)}{2d_c^2}\right\}$$
$$= 2 \exp\left\{-m\delta^2 \frac{\beta^2 R}{2d_c^2 (1-D)}\right\}$$
$$= 2 \exp\left\{-m\delta^2 \frac{\beta^2}{2d_v^2 R(1-D)}\right\},$$

where the final step uses the relation $d_c = R d_v$. ∎

Using Lemma 4 and Proposition 1, we can now prove Corollary 2. Assume that for some pair $(R, D)$ and code $\mathbb{C}$ drawn from the check-regular ensemble, the source encoder is successful. Recall that $F(\mathbb{C})$ is a lower bound on $\log W(D; \mathbb{C})$. Using the fact that $m = Rn$, equation (21) implies that

$$\frac{1}{n}\mathbb{E}[\log W(D; \mathbb{C})] \geq \; := R\frac{d_v!}{2}\left(\frac{D}{d_v}\right)^{d_v} := \alpha(\mathfrak{C}) = R\beta.$$

Moreover, we have

$$\mathbb{P}[\log W(D; \mathbb{C}) \leq (1-\delta)\alpha(\mathbb{C})n] \leq \mathbb{P}[F(\mathbb{C}) \leq (1-\delta)\beta m]$$
$$= \mathbb{P}[F(\mathbb{C}) \leq (1-\delta)\mathbb{E}[F(\mathbb{C})]]$$
$$\leq 2 \exp\left\{-m\delta^2 \frac{\beta^2}{2d_v^2 R(1-D)}\right\}$$
$$= 2\, 2^{-m \, \log_2(e) \, \delta^2 \frac{\beta^2}{2d_v^2 R(1-D)}},$$

using equation (22). Consequently, the hypotheses of Proposition 1 are satisfied with $\beta = \frac{d_v!}{2}\left(\frac{D}{d_v}\right)^{d_v}$, $\alpha(\mathfrak{C}) = R\beta$, $K = 2$, and

$$f(\delta; \beta) := R \, \log_2(e) \left[\delta^2 \frac{\beta^2}{2d_v^2 R(1-D)}\right] =: R\, g(\delta; \beta).$$

Applying Proposition 1, we conclude that

$$R \geq 1 - H(D) + R \max_{\delta \in (0,1)} \min \{g(\delta; \beta), (1-\delta)\beta\} - o(1), \tag{27}$$

thereby establishing the claim of Corollary 2.

## 5 Discussion

We developed a technique for generating lower bounds on the effective rate-distortion function of sparse graph codes. The basic underlying ideas are the source coding analogs of Gallager's [9] classical work on the effective channel capacity of bounded degree codes. Our main result (Theorem 1) provides a generic lower bound on the best possible distortion achievable by any family of rate $R$ codes. The essential object is the excess rate function, corresponding to a certain measure of the average overlap between adjacent codewords. Using this theorem to obtain lower bounds for specific code families requires methods for computing or lower bounding this excess rate term. In this paper, we we illustrated this approach by obtain lower bounds for random ensembles of sparse graph codes, including check-regular ensemble and the bit-check-regular ensembles of LDGM codes. We note that the basic ideas are more generally applicable to other sparse code ensembles, such as LDGM families with prescribed bit and check degree distributions. Moreover, recent work by Kudekar and Urbanke [12] has shown how similar ideas can be used to obtain lower bounds on fixed code sequences, as opposed to the random ensembles considered here.

### Acknowledgments:

This work was partially supported by NSF grants CAREER-CCF-0545862, CCF-0635372 and a Sloan Foundation Fellowship. A preliminary version of portions of this work were presented in part at the Information Theory Workshop, Lake Tahoe, CA in September 2007. We thank Emin Martinian for inspiring discussions.